\documentclass[aps,showpacs,prd,twocolumn]{revtex4}%
\usepackage{graphicx}
\usepackage{amssymb}
\usepackage{amsmath}
\usepackage{color}
\usepackage{float}
\usepackage{accents}
\usepackage{graphicx}
\usepackage{graphicx}
\usepackage{amsfonts}
\usepackage[colorlinks=true, pdfstartview=FitV,
linkcolor=blue,citecolor=blue,urlcolor=blue, breaklinks=true]{hyperref}%
\providecommand{\U}[1]{\protect\rule{.1in}{.1in}}
\begin{document}
\title{Constraining CPT-even and Lorentz-violating nonminimal couplings with the
electron magnetic and electric dipole moments }
\author{Jonas B. Araujo}
\email{jonas.araujo88@gmail.com}
\author{Rodolfo Casana}
\email{rodolfo.casana@gmail.com}
\author{Manoel M. Ferreira Jr}
\email{manojr.ufma@gmail.com}
\affiliation{Departamento de F\'{\i}sica, Universidade Federal do Maranh\~{a}o, Campus
Universit\'{a}rio do Bacanga, S\~{a}o Lu\'{\i}s - MA, 65080-805 - Brazil}

\begin{abstract}
We analyze some dimension-five CPT-even and Lorentz-violating nonminimal
couplings between fermionic and gauge fields in the context of the Dirac
equation. After evaluating the nonrelativistic Hamiltonian, we discuss the
behavior of the terms under discrete symmetries and analyze the implied
effects. We then use the anomalous magnetic dipole moment and electron
electric dipole moment measurements to reach upper bounds of $1$ part in
$10^{20}$ and $10^{24}$ (eV )$^{-1}$, improving the level of restriction on
such couplings by at least 8 orders of magnitude. These upper bounds are also
transferred to the Sun-centered frame by considering the Earth's rotational motion.

\end{abstract}

\maketitle

\section{Introduction}

The existence of the electric dipole of elementary particles may be seen as a
consequence of asymmetric charge distribution along the spin $\mathbf{S}$
direction. Its measurement serves as an important probe for the development of
several distinct theories and new interactions. The electric dipole
interaction is represented by $d ( \boldsymbol{\sigma}\cdot\mathbf{E} ) ,$
with $\mathbf{E}$ being the electric field, $\boldsymbol{\sigma}$ \ the spin
operator, and $d$ the modulus of the electric dipole moment (EDM). An
elementary (or not) particle can only present EDM when both parity ($P$) and
time-reversal ($T$) symmetries are lost, $P ( \boldsymbol{\sigma}%
\cdot\mathbf{E} ) \rightarrow- ( \boldsymbol{\sigma}\cdot\mathbf{E} )$, $T (
\boldsymbol{\sigma}\cdot\mathbf{E}) \rightarrow-( \boldsymbol{\sigma}%
\cdot\mathbf{E}) ,$ so that the presence of EDM is associated with $CP$ and
$T$ violation. In usual electrodynamics, the EDM stems from the Lagrangian
contribution, $id( \bar{\psi}\sigma_{\mu\nu}\gamma_{5}F^{\mu\nu}\psi) $
\cite{EDM1,EDM2,LeptonEDM}, where $\psi$ stands for a Dirac spinor. Elementary
particles really possess a tiny EDM, whose values can be used to constrain the
magnitude of new couplings and theories that induce this kind of physical
behavior \cite{Pospelov}.

Some interesting recent works have been proposed to use the EDM of particles
and atoms as a key factor for constraining the theoretical possibilities of
interaction. A few $CPT$-odd dimension-five interaction terms linear in the
gauge field, $c^{\nu}\bar{\psi}\gamma^{\mu}F_{\mu\nu}\psi,$ $d^{\nu}\bar{\psi
}\gamma^{\mu}\gamma_{5}F_{\mu\nu}\psi,$ $f^{\nu}\bar{\psi}\gamma^{\mu}%
\tilde{F}_{\mu\nu}\psi,$ $g^{\nu}\bar{\psi}\gamma^{\mu}\gamma_{5}\tilde
{F}_{\mu\nu}\psi,$ were analyzed in Ref. \cite{Pospelov}, making the
connection with the EDM generation and Lorentz-violating theories. The
relation of these LV terms with anomalous magnetic moment (MDM) were
considered in Ref. \cite{Stadnik}, which developed an analysis involving the
splitting of the $g$ factors of a fermion and antifermion to constrain some of
them. These $CPT$-odd terms constitute nonminimal couplings between fermions
and photons. The term $g^{\nu}\bar{\psi}\gamma^{\mu}\gamma_{5}\tilde{F}%
_{\mu\nu}\psi,$ for instance, was first proposed in Ref. \cite{NM1} by means
of the nonminimal derivative, $D_{\mu}=\partial_{\mu}+ieA_{\mu}+i\frac
{\lambda}{2}\epsilon_{\mu\lambda\alpha\beta}g^{\lambda}F^{\alpha\beta}$,
defined in the context of the Dirac equation, $(i\gamma^{\mu}D_{\mu}-m)\Psi
=0$, where $g^{\mu}$ can be identified with the Carroll-Field-Jackiw
four-vector, $(k_{AF})^{\mu}=($v$_{0},\mathbf{v),}$ and $\lambda$ is the
coupling constant. This particular coupling was already analyzed in several
respects \cite{NM3,NMmaluf,NMbakke}, including the radiative generation of
$CPT$-odd LV terms \cite{Radio}. See also Ref. \cite{NMABC} and the references therein.

The investigation of Lorentz-symmetry violation is actually a rich line of
research, much developed in the framework of the Standard Model extension
(SME) \cite{Colladay,fermion,CPT,fermion2,KM1,photons1}, whose developments
have scrutinized the Lorentz-violating effects in distinct physical systems
and served to state tight upper bounds on the LV coefficients, including
photon-fermion interactions \cite{Vertex}. Lorentz-violating scenarios are
connected with the breaking of $CPT$ symmetry, although it is known that $CPT$
violation does not necessarily lead to the loss of Lorentz invariance and vice
versa in nonlocal theories \cite{Chaichian}. LV theories are also related to
models containing nonminimal couplings with higher-order derivative terms in
what is called the nonminimal extension of the SME \cite{NMSME}. Alternative
investigations with higher derivatives \cite{HD} and higher-dimension
operators \cite{Reyes} have also been proposed. A recent and broad
investigation about LV effects on the muon MDM was performed in Ref.
\cite{Gomes}, while LV connections with the neutron EDM were reported in Ref.
\cite{Altarev}. The LV contributions that enhance the EDM of charged leptons
stemming from a $CPT$-even term of the SME fermion sector, $d_{\mu\nu}%
\bar{\psi}\gamma_{5}\gamma^{\mu}\psi,$ was developed in Ref. \cite{Haghig},
with the form factor one-loop evaluation.

We have studied a dimension-five $CPT$-even nonminimal coupling in the context
of the Dirac equation \cite{FredeNM1},
\begin{equation}
D_{\mu}=\partial_{\mu}+ieA_{\mu}+\frac{\lambda}{2}(K_{F})_{\mu\nu\alpha\beta
}\gamma^{\nu}F^{\alpha\beta},\label{cov_even}%
\end{equation}
not contained in the broader nonminimal extension of SME. Here, $(K_{F}%
)_{\mu\nu\alpha\beta}$ is \textbf{\ }the $CPT$-even tensor of the SME
electrodynamics. It has the same symmetries of the Riemann tensor and a double
null trace, implying $19$ components \cite{KM1}. Inserted in the Dirac
equation, it provides a nonrelativistic Hamiltonian endowed with contributions
to the EDM and to the MDM, namely $\lambda(\boldsymbol{\sigma}\cdot
\mathbb{\tilde{E})}$, $\lambda(\boldsymbol{\sigma}\cdot\mathbb{\tilde{B})}$,
which rendered upper bounds on the Lorentz-violating parameters as good as
$\lambda\left(  \kappa_{HE}\right)  _{33}\leq10^{-11}(eV)^{-1}$. This
nonminimal coupling radiatively generates the $CPT$-even gauge term of the SME
Lagrangian, $(K_{F})_{\mu\nu\alpha\beta}F^{\mu\nu}F^{\alpha\beta}$
\cite{FredeNM3}. Related studies arguing the generation of topological phases
have been reported as well \cite{Bakke2}.

In the present paper, we propose an axial version of the $CPT$-even nonminimal
coupling considered in Eq. (\ref{cov_even}) in the context of the Dirac
equation. We first access the nonrelativistic regime, evaluating the
associated Hamiltonian from the Dirac's equation. We then analyze the effects
induced on the magnetic dipole and electric dipole moment, using measures of
the electron anomalous MDM and the electron EDM to limit the magnitude of the
nonminimal LV terms at the stringent level of $1$ part in $10^{20}$
(eV)$^{-1}$ and $1$ part in $10^{24}$ (eV)$^{-1}$, respectively.

\section{An axial CPT-even Lorentz-violating nonminimal coupling}

We begin by proposing a quantum electrodynamics where the spinors interact
nonminimally with the electromagnetic field. This is implemented introducing a
dimension-five axial and $CPT$-even nonminimal coupling,
\begin{equation}
D_{\mu}=\partial_{\mu}+ieA_{\mu}+i\frac{\lambda_{A}}{2}(K_{F}) _{\mu\nu
\alpha\beta}\gamma_{5}\gamma^{\nu}F^{\alpha\beta},\label{covader}%
\end{equation}
in the context of the Dirac equation, $(i\gamma^{\mu}D_{\mu}-m)\Psi=0.$ Here,
$(K_{F}) _{\mu\nu\alpha\beta}$ is the $CPT$-even tensor of the SME that can be
written in terms of four $3\times3$ matrices $\kappa_{DE},\kappa_{DB}%
,\kappa_{HE},\kappa_{HB},$ defined in Refs. \cite{KM1} as
\begin{align}
(\kappa_{DE}) _{jk}  &  =-2(K_{F}) _{0j0k},\label{Par1}\\
\text{ }\left(  \kappa_{HB}\right)  _{jk}  &  =\frac{1}{2}\epsilon
_{jpq}\epsilon_{klm}(K_{F}) _{pqlm},\\
\text{ }\left(  \kappa_{DB}\right)  _{jk}  &  =-\left(  \kappa_{HE}\right)
_{kj}=\epsilon_{kpq}(K_{F}) _{0jpq}.\label{Par2}%
\end{align}
The symmetric matrices $\kappa_{DE},\kappa_{HB}$ contain the parity-even
components and possess together $11$ independent components, while
$\kappa_{DB},\kappa_{HE}$ possess no symmetry, having together $8$ components,
representing the parity-odd sector of the tensor $\left(  K_{F}\right)  $.
This classification holds only in the context of a minimal coupling QED. In
the case of a QED with nonminimal interaction involving the tensor $(K_{F}) $,
the parameters $\kappa_{DE},\kappa_{DB},\kappa_{HE},\kappa_{HB}$ could play
distinct roles concerning parity and time reversal, as it appears in Table I.

The Dirac equation can be explicitly written as
\begin{equation}
\left[  i\gamma^{\mu}\partial_{\mu}-e\gamma^{\mu}A_{\mu}-i\frac{\lambda_{A}%
}{2}(K_{F}) _{\mu\nu\alpha\beta}\gamma_{5}\sigma^{\mu\nu}F^{\alpha\beta
}-m\right]  \Psi=0,\label{DiracM1}%
\end{equation}
with the constant $\lambda_{A}$ highlighting the axial character of the
coupling, and
\begin{equation}
\sigma^{\mu\nu}=\frac{i}{2}(\gamma^{\mu}\gamma^{\nu}-\gamma^{\nu}\gamma^{\mu
})=\frac{i}{2}[\gamma^{\mu},\gamma^{\nu}].\label{OP1}%
\end{equation}
Note that $(K_{F})_{\mu\nu\alpha\beta}\gamma_{5}\sigma^{\mu\nu}F^{\alpha\beta
}\Psi$ constitutes a tensor generalization of the usual dipole $\sigma_{\mu
\nu}\gamma_{5}F^{\mu\nu}\Psi$ term. Using the parametrization (\ref{Par1}%
)-(\ref{Par2}), we obtain
\begin{equation}
(K_{F}) _{\mu\nu\alpha\beta}\gamma_{5}\sigma^{\mu\nu}F^{\alpha\beta}%
=2i\Sigma^{j}\left(  \mathbb{E}^{j}-\mathbb{B}^{j}\right)  +2\alpha^{j}\left(
\mathbb{\tilde{E}}^{j}-\mathbb{\tilde{B}}^{j}\right)  ,
\end{equation}
where we have introduced the following rotated fields:
\begin{align}
\ \mathbb{E}^{k}  &  =(\kappa_{DE}) _{kj}E^{j},\text{ \ \ }\mathbb{B}%
^{k}=\left(  \kappa_{DB}\right)  _{kj}B^{j},\\
& \nonumber\\
\mathbb{\tilde{E}}^{k}  &  =\left(  \kappa_{HE}\right)  _{kq}E^{q},\text{
\ }\mathbb{\tilde{B}}^{k}=\left(  \kappa_{HB}\right)  _{kp}B^{p},
\end{align}
with $\kappa_{DE}$, $\kappa_{DB}$, $\kappa_{HE}$, $\kappa_{HB}$, being the
Lorentz-violating matrices. Here, $F_{0j}=E^{j},F_{mn}=\epsilon_{mnp}B_{p}$,
while $\sigma^{0j}=i\alpha^{j},$ $\sigma^{ij}=\epsilon_{ijk}\Sigma^{k},$ and
\[
\alpha^{i}=\left(
\begin{array}
[c]{cc}%
0 & \sigma^{i}\\
\sigma^{i} & 0
\end{array}
\right)  ,\text{ \ \ \ \ }\Sigma^{k}=\left(
\begin{array}
[c]{cc}%
\sigma^{k} & 0\\
0 & \sigma^{k}%
\end{array}
\right)  ,
\]
and $\mathbf{\sigma}=(\sigma_{x},\sigma_{y},\sigma_{z})$ are the Pauli
matrices. In the momentum coordinates, $i\partial_{\mu}\rightarrow p_{\mu},$
the corresponding Dirac equation is
\begin{equation}
i\partial_{t}\Psi=\left[  \boldsymbol{\alpha}\cdot\boldsymbol{\pi}%
+eA_{0}+m\gamma^{0}-\lambda_{A}\gamma^{0}\Sigma^{k}\mathbb{Z}^{k}+i\lambda
_{A}\gamma^{k}\mathbb{\tilde{Z}}^{k}\right]  \Psi,\label{DiracM3}%
\end{equation}
with $\boldsymbol{\pi}=\boldsymbol{p}-e\boldsymbol{A}$ being the canonical
momentum and
\begin{equation}
\mathbb{Z}=\left(  \mathbb{E}-\mathbb{B}\right)  ,~\ \mathbb{\tilde{Z}%
}=\left(  \mathbb{\tilde{E}}-\mathbb{\tilde{B}}\right)  .
\end{equation}
We point out that the presence of the factor $\gamma^{0},$ multiplying the
term $\Sigma^{k}\mathbb{Z}^{k},$ implies an effective contribution to the
energy of the system, evading the Schiff's theorem and allowing us to use the
electron EDM data to constrain the magnitude of this axial coupling.

In order to investigate the role played by this nonminimal coupling, we should
evaluate the nonrelativistic limit of the Dirac equation. Writing the
spinor\thinspace$\ \Psi$\ in terms of small $\left(  \chi\right)  $\ and large
$\left(  \phi\right)  $\ two-spinors,
\begin{equation}
\Psi=\left(
\begin{array}
[c]{c}%
\phi\\
\chi
\end{array}
\right)  ,
\end{equation}
the Dirac equation (\ref{DiracM3}) leads to two $2$-component equations,%
\begin{align}
\left[  E-eA_{0}-m+\lambda_{A}\sigma^{j}\mathbb{Z}^{j}\right]  \phi-\left[
\boldsymbol{\sigma}\cdot\boldsymbol{\pi}+i\lambda_{A}\sigma^{j}\mathbb{\tilde
{Z}}^{j}\right]  \chi &  =0,\\
& \nonumber\\[-0.2cm]
\left[  \boldsymbol{\sigma}\cdot\boldsymbol{\pi}-i\lambda_{A}\sigma
^{j}\mathbb{\tilde{Z}}^{j}\right]  \phi-\left[  E-eA_{0}+m-\lambda_{A}%
\sigma^{j}\mathbb{Z}^{j}\right]  \chi &  =0.
\end{align}

At first order in the Lorentz violating parameters, the following
nonrelativistic Hamiltonian is achieved for the case of uniform fields:
\begin{align}
H_{A}  &  =\frac{1}{2m}\left[  (\boldsymbol{p}-e\boldsymbol{A)}^{2}-e\left(
\boldsymbol{\sigma\cdot B}\right)  \right]  +eA_{0}-\lambda_{A}%
(\boldsymbol{\sigma}\cdot\mathbb{Z)}\nonumber\\
& \label{HNR3}\\
&  +\frac{\lambda_{A}}{m}\mathbb{\tilde{Z}}\cdot(\boldsymbol{\sigma}%
\times\mathbf{p})-\frac{e\lambda_{A}}{m}\mathbb{\tilde{Z}}\cdot(\mathbf{\sigma
}\times\mathbf{A}).\nonumber
\end{align}

Concerning the new effects induced by this Hamiltonian, we are particularly
interested in the terms that lead to corrections to the anomalous magnetic
moment, $\lambda_{A}(\boldsymbol{\sigma}\cdot\mathbb{B)},$ and to the electric
dipole moment of the electron, $\lambda_{A}(\boldsymbol{\sigma}\cdot
\mathbb{E)}$. \ We also note that the term $\mathbb{E}\cdot(\boldsymbol{\sigma
}\times\mathbf{p})$ is a generalization of the Rashba coupling term, while
$\boldsymbol{\sigma}\cdot\mathbb{E}$ also generates a kind of electric Zeeman
effect, in the total absence of a magnetic field.

A parallel can now be traced to the nonminimal coupling of Ref.
\cite{FredeNM1}, whose Dirac equation,
\begin{equation}
\left[  i\gamma^{\mu}\partial_{\mu}-e\gamma^{\mu}A_{\mu}+\frac{\lambda}%
{2}(K_{F}) _{\mu\nu\alpha\beta}\sigma^{\mu\nu}F^{\alpha\beta}-m\right]
\Psi=0,\label{DiracM7}%
\end{equation}
\begin{equation}
i\partial_{t}\Psi=\left[  \boldsymbol{\alpha}\cdot\boldsymbol{\pi}%
+eA_{0}+i\lambda\gamma^{i}\mathbb{Z}^{i}+\lambda\gamma^{0}\Sigma
^{k}\mathbb{\tilde{Z}}^{k}+m\gamma^{0}\right]  \Psi,\label{DiracM8}%
\end{equation}
yields the following LV corrections to the nonrelativistic Hamiltonian:
\begin{equation}
H_{LV}=-\lambda(\boldsymbol{\sigma}\cdot\mathbb{\tilde{Z})}-\frac{\lambda}%
{m}\mathbb{Z}\cdot(\boldsymbol{\sigma}\times\mathbf{p})+\frac{e\lambda}%
{m}\mathbb{Z}\cdot(\mathbf{\sigma}\times\mathbf{A}),\label{HLV4}%
\end{equation}
which reveals an EDM\ term even in the absence of the $\gamma_{5}$ matrix in
the nonminimal coupling of Eq. (\ref{DiracM7}).

We are ready to discuss the behavior of the modified Dirac equations
(\ref{DiracM3}) and (\ref{DiracM8}) under the discrete symmetries $C,P,T$.
While the LV parameters of Eq. (\ref{DiracM8}) obey the original
classification of the $3\times3$ matrices \cite{KM1} under $P$ and $T$, the
components of the axial coupling follow reversed behavior under such
operators. Table I displays the response under the $C,P,T$ operators of the
axial coupling parameters of Hamiltonian (\ref{HNR3}), $\lambda_{A}%
(\kappa_{DE}),$ $\lambda_{A}\left(  \kappa_{DB}\right)  ,$ $\lambda_{A}%
(\kappa_{HE}),$ $\lambda_{A}(\kappa_{HB}),$ and the coupling parameters of
Hamiltonian (\ref{HLV4}), $\lambda(\kappa_{DE}),$ $\lambda\left(  \kappa
_{DB}\right)  ,$ $\lambda(\kappa_{HE}),$ $\lambda(\kappa_{HB}).$ We notice
that the elements $\lambda_{A}(\kappa_{DE})$, $\lambda_{A}(\kappa_{HB})$ and
$\lambda_{A}(\kappa_{DB})$, $\lambda_{A}(\kappa_{HE})$ are now $P$ odd and $P
$ even, respectively, instead of inheriting the usual behavior of the matrices
$(\kappa_{DE}),(\kappa_{HB}),(\kappa_{DB}),(\kappa_{HE}).$ We can also observe
that the axial term, $\lambda_{A}\Sigma^{i}E^{i},$ that yields the
nonrelativistic interaction $\lambda_{A}(\sigma\cdot E),$ is $P$ odd and $T$
odd, compatible with the EDM character. The same holds for the term
$\lambda\Sigma^{k}\tilde{E}^{k}$ of Hamiltonian (\ref{HLV4}).

\begin{table}[h]
\centering
\begin{tabular}
[c]{|l|l|l|l|l|}\hline
$ANM$ & $\lambda_{A}(\kappa_{DE})$ & $\lambda_{A}\left(  \kappa_{DB}\right)  $
& \ $\lambda_{A}(\kappa_{HE})$ & $\lambda_{A}(\kappa_{HB})$\\\hline
$\ C$ & $\ \ \ \ +$ & $\ \ \ \ +$ & $\ \ \ \ +$ & $\ \ \ \ +$\\\hline
$\ P$ & $\ \ \ \ -$ & $\ \ \ \ +$ & $\ \ \ \ +$ & $\ \ \ \ -$\\\hline
$\ T$ & $\ \ \ \ -$ & $\ \ \ \ +$ & $\ \ \ \ +$ & $\ \ \ \ -$\\\hline
$CPT$ & $\ \ \ \ +$ & $\ \ \ \ +$ & $\ \ \ \ +$ & \ \ \ \ $+$\\\hline
&  &  &  & \\\hline
$NM$ & $\lambda(\kappa_{DE})$ & $\lambda\left(  \kappa_{DB}\right)  $ &
\ $\lambda(\kappa_{HE})$ & $\lambda(\kappa_{HB})$\\\hline
$\ \ C$ & $\ \ \ \ +$ & $\ \ \ \ +$ & $\ \ \ \ +$ & $\ \ \ \ +$\\\hline
$\ P$ & $\ \ \ \ +$ & $\ \ \ \ -$ & $\ \ \ \ -$ & $\ \ \ \ +$\\\hline
$\ T$ & $\ \ \ \ +$ & $\ \ \ \ -$ & $\ \ \ \ -$ & $\ \ \ \ +$\\\hline
$CPT$ & $\ \ \ \ +$ & $\ \ \ \ +$ & $\ \ \ \ +$ & \ \ \ \ $+$\\\hline
\end{tabular}
\caption{Complete classification under $C,P,T$ for the coefficients of the
axial nonminimal coupling (ANM) and usual nonmimal coupling (NM). }%
\label{Table1}%
\end{table}

\section{Tree-level magnetic and electric dipole moments}

An interesting feature of the Hamiltonian (\ref{HNR3}) is that the term
$\lambda_{A}\boldsymbol{\sigma}\cdot\left(  \mathbb{E}-\mathbb{B}\right)  $ is
able to generate a magnetic moment and an electric dipole moment.
Investigations about LV effects on the electron anomalous MDM were developed
in Refs. \cite{Carone}.

The electron magnetic moment is $\boldsymbol{\mu}=-g\mu_{B}\boldsymbol{S}$,
where $\mu_{B}=e/2m$, $\boldsymbol{S}$ is the spin operator and $g=2$ is the
gyromagnetic factor. The anomalous magnetic moment of the electron is given by
$g=2(1+a),$ with $a=\alpha/2\pi\simeq0.00116$ representing the deviation in
relation to the usual case. Its most precise calculation is found in Ref.
\cite{anomalousMP}. The magnetic interaction is $H^{\prime}=\mu_{B}%
(1+a)\left(  \boldsymbol{\sigma}\cdot\mathbf{B}\right)  $. In accordance with
very precise measurements \cite{Gabrielse}, the experimental error on the
electron MDM is at the level of $2.8$\ parts in $10^{13},$\ that is, $\Delta
a\leq2.8\times10^{-13}.$\ In our case, the Hamiltonian (\ref{HNR3}) provides
tree-level LV\ contributions to the usual $g=2$\ gyromagnetic factor, which
cannot be larger than $\Delta a.$ The total magnetic interaction in Eq.
(\ref{HNR3}) is $\mu_{B}\left(  \boldsymbol{\sigma\cdot B}\right)
+\lambda_{A}\left(  \boldsymbol{\sigma}\cdot\mathbb{B}\right)  .$ For the
magnetic field along the $z$ axis, $\mathbf{B=}B_{0}\hat{z}$, and a
spin-polarized configuration in the $z$ axis, this interaction assumes the
form
\begin{equation}
\mu_{B}\left[  1+\frac{2m}{e}\lambda_{A}\left(  \kappa_{DB}\right)
_{33}\right]  \left(  \sigma_{z}B_{0}\right)  ,
\end{equation}
where $\frac{2m}{e}\lambda_{A}\left(  \kappa_{DB}\right)  _{33}$ stands for
the tree-level LV correction that should be smaller than $\Delta a$, so that
\begin{equation}
\left\vert \lambda_{A}\left(  \kappa_{DB}\right)  _{33}\right\vert
\leq2.3\times10^{-20}\,(\mbox{eV})^{-1},\label{MDb1}%
\end{equation}
represents an improvement by a factor $\simeq10^{10}$ on the strength of the
corresponding bound of Ref. \cite{FredeNM1}. Alternatively, if we use the
analysis of Ref. \cite{Stadnik}, based on the splitting of the $g$ factor of
electron and positron, we conclude that $\lambda_{A}\left(  \kappa
_{DB}\right)  _{33}\leq2.3\times10^{-12}\mu_{B}$, leading to
\begin{equation}
\left\vert \lambda_{A}\left(  \kappa_{DB}\right)  _{33}\right\vert
\leq1.9\times10^{-19}\,(\mbox{eV})^{-1}.\label{bound1}%
\end{equation}

We should now discuss the EDM\ term, $\lambda_{A}(\boldsymbol{\sigma}%
\cdot\mathbb{E)}$, which can be written as
\begin{equation}
\lambda_{A}\left(  \boldsymbol{\sigma}\cdot\mathbb{E}\right)  =\lambda
_{A}\frac{(\kappa_{DE}) _{ii}}{3}\left(  \boldsymbol{\sigma}\cdot
\mathbf{E}\right)  .
\end{equation}
The strongest limitations to be achieved involve the electron EDM,
$\mathbf{d}_{e},$ that is the minor known one. The magnitude of $\mathbf{d}%
_{e}$ has been constrained with increasing precision
\cite{EDM1,Measure1,Measure2}, reaching the level $\left\vert \mathbf{d}%
_{e}\right\vert \leq1.1\times10^{-29}$e.m or $\left\vert \mathbf{d}%
_{e}\right\vert \leq4.7\times10^{-24}$ (eV)$^{-1}$. Very recent experiments
\cite{Baron} improved this limit as $\left\vert \mathbf{d}_{e}\right\vert
\leq8.7\times10^{-31}$e.m , or
\begin{equation}
\left\vert \mathbf{d}_{e}\right\vert \leq3.8\times10^{-25}(\mbox{eV})^{-1}.
\end{equation}
Considering this experimental measure, we attain the following upper bound:
\begin{equation}
\left\vert \lambda_{A}(\kappa_{DE}) _{ii}\right\vert \leq1.1\times
10^{-24}(\mbox{eV})^{-1},\label{bound2}%
\end{equation}
surpassing the best magnetic bound (\ref{MDb1}) by the factor $10^{4}$. We
remark that this limit is at least $8$ orders of magnitude better than the
bounds first attained in the analogue nonminimal coupling of Ref.
\cite{FredeNM1}.

We know that the nonminimal coupling of Ref. \cite{FredeNM1} yields an EDM
term even without containing a $\gamma_{5}$ matrix. Therefore, we can also
improve the upper bounds on some of its components by the same procedure. We
begin with the term $\lambda(\boldsymbol{\sigma}\cdot\mathbb{\tilde{E})}$ of
Eq. (\ref{HLV4}) using the electron EDM. In this case, as the matrix
$\kappa_{HE}$ is traceless, we should choose a particular direction for the
electric field, $\mathbf{E=}E_{0}\hat{x}$, which yields
\begin{equation}
\lambda(\boldsymbol{\sigma}\cdot\mathbb{\tilde{E})}=\lambda\left(  \kappa
_{HE}\right)  _{11}\left(  \boldsymbol{\sigma}_{x}\mathbf{E}_{0}\right)  ,
\end{equation}
implying
\begin{equation}
\lambda\left(  \kappa_{HE}\right)  _{11}\leq3.8\times10^{-25}(eV)^{-1}%
,\label{bound3}%
\end{equation}
which represents an improvement of the corresponding previous bound of
\cite{FredeNM1} by a factor $\sim10^{8}$.

As for the term $\lambda(\boldsymbol{\sigma}\cdot\mathbb{\tilde{B})}$ of Eq.
(\ref{HLV4}), we use the anomalous electron MDM measures to improve the
previous bound by a factor $\sim10^{10},$ that is,
\begin{equation}
\lambda\left(  \kappa_{HB}\right)  _{33}\leq2.3\times10^{-20}%
\,(\mbox{eV})^{-1}.\label{bound4}%
\end{equation}
In these evaluations, we have used natural units: $m_{e}=5.11\times
10^{5}\ \mbox{eV}$, $e=\sqrt{1/137}$, and $1$m$=5.06\times10^{6}%
\,(\mbox{eV})^{-1}.$

\section{Sidereal Variations}

Strictly speaking, neither the Earth nor the Sun is an ideal inertial
reference frame (RF). Nevertheless, the latter is closer to being one, as its
rotation period, around the center of the galaxy, is about 230 million years.
Since the Lorentz-violating (LV) tensors are constant for an inertial RF, the
time dependence of their components is expected in experiments performed on
the Earth, exhibiting a periodic variation associated with the Earth's
rotation time $\left(  1/\Omega\right)  $. A reasonable choice for an inertial
RF is the Sun, with the $z$ axis matching the direction of the Earth's
rotation axis and the $x$ axis pointing from the Earth's center to the Sun on
the vernal equinox in 2000. For more details, see Refs. \cite{Sideral}. In
experiments up to a few weeks long, it is possible to neglect the Earth's
motion around the Sun, so that the transformation on the LV tensors is a mere
rotation, due to the Earth's rotation around its own axis. According to this,
a $3$-component rank-1 tensor, on the spinning Earth's RF, transforms as
\begin{equation}
V_{i}^{\text{Lab}}=\mathcal{R}_{ik}V_{k}^{\text{Sun}}\ ,
\end{equation}
where
\begin{equation}
\mathcal{R}_{ij}=%
\begin{pmatrix}
\cos\chi\cos\Omega t & \cos\chi\sin\Omega t & -\sin\chi\\
-\sin\Omega t & \cos\Omega t & 0\\
\sin\chi\cos\Omega t & \sin\chi\sin\Omega t & \cos\chi
\end{pmatrix}
,
\end{equation}
in which $\chi$ is the colatitude of the lab and $\Omega=2\pi/23$h$\,56$ min
is the Earth's rotation angular velocity. A rank-2 tensor transforms according
to the rule
\begin{equation}
A_{ij}^{\text{Lab}}=\mathcal{R}_{ik}\mathcal{R}_{jl}A_{kl}^{\text{Sun}}\ .
\end{equation}
In the literature \cite{Sideral}, the Earth-based frame, where the laboratory
is located, has axis $x$, $y$, $z$, while the Sun-centered frame has $X$, $Y$,
$Z$ as axis. Hence, $A_{kl}^{\text{Sun}}\equiv A_{kl}^{(X,Y,Z)}$,
$\ A_{ij}^{\text{Lab}}\equiv A_{ij}^{(x,y,z)}.$ Applying this transformation
to the $33$-component of the matrix, we obtain
\begin{align}
(A)_{33}^{\text{Lab}} &  =(\sin^{2}\chi\cos^{2}\Omega t)A_{11}^{\text{Sun}%
}+(\sin\chi^{2}\cos\Omega t\sin\Omega t)A_{12}^{\text{Sun}}\nonumber\\
&  +(\sin\chi\cos\chi\cos\Omega t)A_{13}^{\text{Sun}}+(\sin^{2}\chi\sin\Omega
t\cos\Omega t)A_{21}^{\text{Sun}}\nonumber\\
&  +(\sin^{2}\chi\sin^{2}\Omega t)A_{22}^{\text{Sun}}+(\sin\Omega t\sin
\chi\cos\chi)A_{23}^{\text{Sun}}\nonumber\\
&  +(\cos\Omega t\cos\chi\sin\chi)A_{31}^{\text{Sun}}+(\sin\Omega t\cos
\chi\sin\chi)A_{32}^{\text{Sun}}\nonumber\\
&  +\cos^{2}\chi A_{33}^{\text{Sun}}\ ,
\end{align}
whose time average leads simply to%
\begin{equation}
\left\langle A_{zz}^{\text{Lab}}\right\rangle =\frac{1}{2}\sin^{2}\chi
A_{11}^{\text{Sun}}+\frac{1}{2}\sin^{2}\chi A_{22}^{\text{Sun}}+\cos^{2}\chi
A_{33}^{\text{Sun}}\ .
\end{equation}

With these transformations, it is possible to write the upper bounds here
attained in terms of combinations of the Sun-based frame ones. The bound
(\ref{MDb1}) yields
\begin{align}
&  \left\vert \frac{1}{2}(\lambda_{A}\kappa_{DB})_{XX}^{\text{Sun}}\sin
^{2}\chi+\frac{1}{2}(\lambda_{A}\kappa_{DB})_{YY}^{\text{Sun}}\sin^{2}%
\chi\right. \nonumber\\
&  \left.  ~~+(\lambda_{A}\kappa_{DB})_{ZZ}^{\text{Sun}}\cos^{2}\chi\frac{{}%
}{{}}\right\vert \leq2.3\times10^{-20}\,(\text{eV})^{-1}\ .
\end{align}

As the trace is invariant under rotation,\ the bound (\ref{bound2}) reads as
\begin{equation}
|\langle\lambda_{A}(\kappa_{DE})_{ii}^{\text{Lab}}\rangle|=|\langle\lambda
_{A}(\kappa_{DE})_{ii}^{\text{Sun}}\rangle|\leq\ 1.1\times10^{-24}%
(\text{eV})^{-1}\ .
\end{equation}

Finally, the bounds (\ref{bound3}) and (\ref{bound4}) then become
\begin{align}
&  \left\vert \frac{1}{2}(\lambda\kappa_{HE})_{XX}^{\text{Sun}}\cos^{2}%
\chi+\frac{1}{2}(\lambda\kappa_{HE})_{YY}^{\text{Sun}}\cos^{2}\chi\right.
\nonumber\\
&  \left.  ~~+(\lambda\kappa_{HE})_{ZZ}^{\text{Sun}}\sin^{2}\chi\frac{{}}{{}%
}\right\vert \leq3.8\times10^{-25}\,(\text{eV})^{-1}\ ,
\end{align}%
\begin{align}
&  \left\vert \frac{1}{2}(\lambda\kappa_{HB})_{XX}^{\text{Sun}}\sin^{2}%
\chi+\frac{1}{2}(\lambda\kappa_{HB})_{YY}^{\text{Sun}}\sin^{2}\chi\right.
\nonumber\\
&  \left.  ~~~+(\lambda\kappa_{HB})_{ZZ}^{\text{Sun}}\cos^{2}\chi\frac{{}}{{}%
}\right\vert \leq2.3\times10^{-20}\,(\text{eV})^{-1}\ .
\end{align}
For attaining a clearer scenario of constraining on the Sun-based components,
it is necessary to know the colatitude $\chi$\ and details of the experimental
device, as the electric alignment.

\section{Conclusions}

We have proposed an axial $CPT$-even, dimension-five, and Lorentz-violating
nonminimal coupling between fermionic and gauge fields, involving the tensor
$(K_{F}) _{\mu\nu\alpha\beta}$ of the gauge sector of the SME, in the context
of the Dirac equation. The nonrelativistic Hamiltonian was carried out,
revealing corrections to the anomalous magnetic moment and to electron EDM.
\ Effects of these terms on the electron EDM and on the anomalous magnetic
moment of electrons and positrons have yielded upper bounds as tight as
$\lambda_{A}(\kappa_{DE}) _{ii}\leq1.1\times10^{-24}\,(\mbox{eV})^{-1}$ and
$\left\vert \lambda_{A}\left(  \kappa_{DB}\right)  _{33}\right\vert
\leq2.3\times10^{-20}\,(\mbox{eV}) ^{-1},$ respectively, which can be also
expressed in terms of the Sun-based frame coefficients. Using the same
procedure, we have shown that it is possible to improve the previous bounds on
the nonminimal coupling of Ref. \cite{FredeNM1} by the factors $10^{8}$ and
$10^{10}$ .

We remark that the $CPT$-even nonminimal coupling proposed in Eq.
(\ref{DiracM1}) evades the Schiff's theorem, implying physical effects in the
energy of the system, $\Delta U=-\mathbf{d}_{e}\cdot\mathbf{E,}$ as explained
in Ref. \cite{LeptonEDM}, which allows to use the electron EDM to attain the
tightest upper bounds on the nonminimal LV\ parameters.\textbf{\ }Such
evasion, however, is not fulfilled by Dirac particles interacting only via
electromagnetic $CPT$-odd nonminimal couplings, requiring the use of the EDM
of composite particles, as discussed in Ref. \cite{Pospelov}. As composite
particles have larger EDM, it leads to weaker upper bounds on the quark sector
by at least a factor $10^{4}$ (when compared with the ones attained with
electron EDM).

The bounds found on the axial coupling coefficients, $\lambda_{A}\left(
\kappa_{DB}\right)  ,$ $\lambda_{A}(\kappa_{DE}) ,$ should not be confused
with the upper bounds on the parameters $\lambda\left(  \kappa_{HB}\right)  ,$
$\lambda\left(  \kappa_{HE}\right)  $ of the first $CPT$-even nonminimal
coupling \cite{FredeNM1}. They constitute restrictions on different but
analogue interactions, both restrictions being distinct from the known upper
bounds on the $CPT$-even parameters of the SME \cite{KM1}.

\begin{acknowledgments}
The authors are grateful to CNPq, CAPES and FAPEMA (Brazilian research
agencies) for invaluable financial support.
\end{acknowledgments}

\end{document}